\documentclass[journal]{IEEEtran}
\usepackage{amsfonts}

\usepackage{amsthm}
\usepackage[fleqn]{amsmath}

\ifCLASSINFOpdf
   \usepackage[pdftex]{graphicx}
\else
   \usepackage[dvips]{graphicx}
   \graphicspath{{../eps/}}
\fi
\usepackage[tight,footnotesize]{subfigure}
\begin{document}
%
\title{Generalized Secure Transmission Protocol for Flexible Load-Balance Control with Cooperative Relays in Two-Hop Wireless Networks}
%
%
%

\author{\IEEEauthorblockN{Yulong Shen\IEEEauthorrefmark{1}\IEEEauthorrefmark{3},
Xiaohong Jiang\IEEEauthorrefmark{2} and Jianfeng
Ma\IEEEauthorrefmark{1}}\\
\IEEEauthorblockA{\IEEEauthorrefmark{1}School of Computer Science
and Technology, Xidian University, China}\\
\IEEEauthorblockA{\IEEEauthorrefmark{2}School of Systems Information
Science, Future University Hakodate, Japan}\\
\IEEEauthorblockA{\IEEEauthorrefmark{3}Email:ylshen@mail.xidian.edu.cn
} }

\maketitle

\begin{abstract}

This work considers secure transmission protocol for flexible
load-balance control in two-hop relay wireless networks without the
information of both eavesdropper channels and locations. The
available secure transmission protocols via relay cooperation in
physical layer secrecy framework cannot provide a flexible
load-balance control, which may significantly limit their
application scopes. This paper extends the conventional works  and
proposes a general transmission protocol with considering
load-balance control, in which the relay is randomly selected from
the first $k$ preferable assistant relays located in the circle area
with the radius $r$ and the center at the middle between source and
destination (2HR-($r,k$) for short). This protocol covers the
available works as special cases, like ones with the optimal relay
selection ($r=\infty$, $k=1$) and with the random relay selection
($r=\infty$, $k = n$ i.e. the number of system nodes) in the case of
equal path-loss, ones with relay selected from relay selection
region ($r \in (0, \infty), k = 1$) in the case of
distance-dependent path-loss. The theoretic analysis is further
provided to determine the maximum number of eavesdroppers one
network can tolerate to ensure a desired performance in terms of the
secrecy outage probability and transmission outage probability. The
analysis results also show the proposed protocol can balance load
distributed among the relays by a proper setting of $r$ and $k$
under the premise of specified secure and reliable requirements.

\end{abstract}



%
\IEEEpeerreviewmaketitle

\section{Introduction}

Wireless networks have the promising applications of in many
important scenarios (like battlefield networks, emergency networks,
disaster recovery networks). However, Due to the energy constrained
and broadcast properties, the consideration of secrecy and lifetime
optimization in such networks is of great importance for ensuring
the high transmission efficiency and confidentiality requirements of
these applications. Two-hop wireless networks, as a building block
for large multi-hop network system, have been a class of basic and
important networking scenarios \cite{IEEEhowto:Sathya}. The analysis
and design of transmission protocol in basic two-hop relay networks
serves as the foundation for secure information exchange of general
multi-hop network system.

For the lifetime optimization, an uneven use of the nodes may cause
some nodes die much earlier, thus creating holes in the network, or
worse, leaving the network disconnected, which is critical in
military or emergency networks. For this problem, a lot of protocols
were proposed to balance the traffic across the various relay nodes
and avoids overloading any relay node in various wireless networks,
especially energy constrained wireless environments (like wireless
sensor networks) [7-16](see Section V for related works). We notice
there is tradeoff between the load-balance capacity and transmission
efficiency and still no approaches can flexibly control it.
Regarding the secrecy, the traditional cryptographic approach can
provide a standard information security. However, the everlasting
secrecy can not be achieved by such approach, because the adversary
can record the transmitted messages and try any way to break them
\cite{IEEEhowto:Talbot}. Especially, recent advances in
high-performance computation (e.g. quantum computing) further
complicate acquiring long-lasting security via cryptographic
approaches \cite{IEEEhowto:Joux}. This motivates the consideration
of signaling scheme in physical layer secrecy framework to provide a
strong form of security, where a degraded signal at an eavesdropper
is always ensured such that the original data can be hardly
recovered regardless of how the signal is processed at the
eavesdropper
\cite{IEEEhowto:Wyner}\cite{IEEEhowto:Vasudevan}\cite{IEEEhowto:Koyluoglu}.

The secure and reliable transmission in physical layer secrecy
framework for two-hop relay wireless networks has been studied and a
lot of secure transmission protocols were proposed in [17-28](see
Section V for related works). These works mainly focus on the
maximum secrecy capacity and minimum energy consumption, in which
the system node with the best link condition to source and
destination is selected as information relay. These protocols are
attractive in the sense that provides very effective resistance
against eavesdroppers. However, since the channel state is
relatively constant during a fixed time period, some relay nodes
with good link conditions always prefer to relay packages, which
results in a severe load-balance problem and a quick node energy
depletion. Such, these protocol is not suitable for energy-limited
wireless networks (like wireless sensor networks). In order to
realize load-balance, Y. Shen et al. further proposed a random relay
selection protocol \cite{IEEEhowto:Shen1}\cite{IEEEhowto:Shen2}, in
which the relay node is random selected from the system nodes.
However, this protocol has lower transmission efficiency. Such, it
is only suitable for large scale wireless network environment with
stringent energy consumption constraint.

In summary, the available secure transmission protocols cannot
provide a flexible load-balance control, which may significantly
limit their application scopes. This paper extends conventional
secure cooperative transmission protocols to a general case to
enable the load-balance to be flexibly controlled in the two-hop
relay wireless networks without the knowledge of eavesdropper
channels and locations. The main contributions of this paper are as
follows:

\begin{itemize}

\item
This paper proposes a new transmission protocol 2HR-($r,k$) for
two-hop relay wireless network without the knowledge eavesdropper
channels and locations, where the relay is randomly selected from
the first $k$ preferable assistant relays located in the circle area
with the radius $r$ and the center at the middle between source and
destination. This protocol is general protocol, and can flexibly
control the tradeoff between the load-balance among relays and the
transmission efficiency by a proper setting of $k$ and $r$ under the
premise of specified secure and reliable requirements.

\item
In case that the path-loss is identical between all pairs of nodes,
theoretic analysis of 2HR-($r,k$) protocol is provided to determine
the corresponding exact results on the number of eavesdroppers one
network can tolerate to satisfy a specified requirement and shows
that the 2HR-($r,k$) protocol covers all the available secure
transmission protocols as special cases, like ones with the optimal
relay selection ($r=\infty$, $k=1$)
\cite{IEEEhowto:Goeckel1}\cite{IEEEhowto:Goeckel2}\cite{IEEEhowto:Vasudevan2}\cite{IEEEhowto:Shen1}
and with the random relay selection ($d=\infty$, $k = n$ i.e. the
number of system nodes)\cite{IEEEhowto:Shen1}\cite{IEEEhowto:Shen2}.

\item
In case that the path-loss between each pair of nodes also depends
on the distance between them, a coordinate system is presented and
the theoretic analysis of 2HR-($r,k$) protocol is provided to
determine the corresponding exact results on the number of
eavesdroppers one network can tolerate to satisfy a specified
requirement and shows that the 2HR-($r,k$) protocol covers all the
available secure transmission protocols as special cases, like ones
with relay selected from relay selection region ($r \in (0, \infty),
k = 1$)\cite{IEEEhowto:Shen2}.

\end{itemize}

The remainder of this paper is organized as follows. Section II
presents system models and the 2HR-($r,k$) protocol. Section III
presents the theoretic analysis in case of equal path-loss between
all node pairs. Section IV presents the theoretic analysis in case
that path-loss between each node pair also depends on their relative
locations. Section V is related works and Section VI concludes this
paper.

\section{System Models and 2HR-($r,k$) Protocol}

\subsection{Network Model}

A Two-hop wireless network scenario is considered where a source
node $S$ wishes to communicate securely with its destination node
$D$ with the help of multiple relay nodes $R_1$, $R_2$, $\cdots$,
$R_n$. Also present in the environment are $m$ eavesdroppers $E_1$,
$E_2$, $\cdots$, $E_m$ without knowledge of channels and locations.
The relay nodes and eavesdroppers are independent and also uniformly
distributed in the network, as illustrated in Fig.1. Our goal here
is to design a general protocol to ensure the secure and reliable
information transmission from source $S$ to destination $D$ and
provide flexible load-balance control among the relays.

\begin{figure}[!t]
\centering
\includegraphics[width=2.5in]{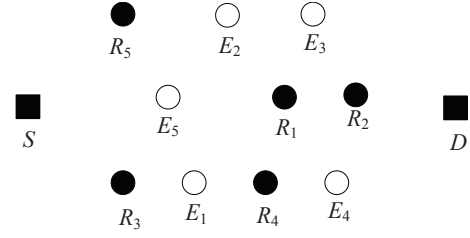}
\DeclareGraphicsExtensions. \caption{System scenario: Source $S$
wishes to communicate securely with destination $D$ with the
assistance of finite relays $R_1$, $R_2$, $\cdots$, $R_{n}$ ($n$=5
in the figure) in the presence of passive eavesdroppers $E_1$,
$E_2$, $\cdots$, $E_{m}$ ($m$=5 in the figure). Cooperative relay
scheme is used in the two-hop transmission.} \label{System scenario}
\end{figure}

\subsection{Transmission Model}

Consider the transmission from a transmitter $A$ to a receiver $B$,
and denote the $i^{th}$ symbol transmitted by node $A$ by
$x_i^{\left(A\right)}$. We assume that all nodes transmit with the
same power $E_s$ and path-loss between all pairs of nodes is
independent. We denote the frequency-nonselective multi-path fading
from $A$ to $B$ by $h_{A,B}$. Under the condition that all nodes in
a group of nodes, $\mathcal {R}$, are generating noises, the
$i^{th}$ signal received at node $B$ from node $A$, denoted by
$y_i^{\left(B\right)}$, is determined as:

$$y_i^{\left(B\right)}=\frac{h_{A,B}}{d_{A,B}^{\alpha / 2}} \sqrt{E_s}x_i^{\left(A\right)} +
\sum_{A_j \in \mathcal {R}} \frac{h_{A_j,B}}{d_{A_j,B}^{\alpha / 2}}
\sqrt{E_s}x_i^{\left(A_j\right)} + n_i^{\left(B\right)}$$

where $d_{A,B}$ is the distance between node $A$ and $B$, $\alpha
\geq 2$ is the path-loss exponent, $\left|h_{A,B}\right|^2$ is
exponentially distributed and without loss of generality, we assume
that $E{\left[\left|h_{A,B}\right|^2\right]}=1$. The noise
$n_i^{\left(B\right)}$ at receiver $B$ is assumed to be i.i.d
complex Gaussian random variables with mean $N_0$. The SINR
$C_{A,B}$ from $A$ to $B$ is then given by

$$C_{A,B}=\frac{E_s\left|h_{A,B}\right|^2 d_{A,B}^{- \alpha}}{\sum_{A_j \in \mathcal
{R}}E_s{\left|h_{A_j,B}\right|^2 d_{A_j,B}^{- \alpha}}+N_0/2}$$

For a legitimate node and an eavesdropper, we use two separate SINR
thresholds $\gamma_R$ and $\gamma_E$ to define the minimum SINR
required to recover the transmitted messages for legitimate node and
eavesdropper, respectively. Therefore, a system node (the selected
relay or destination) is able to decode a packet if and only if its
received SINR is greater than $\gamma_R$, whereas each eavesdropper
try to achieve target SINR $\gamma_E$ to recover the transmitted
message. However, from an information-theoretic perspective, we can
map to a secrecy rate formulation $R \geq \frac{1}{2}
\log(1+\gamma_R) - \frac{1}{2} \log(1+\gamma_E)$
\cite{IEEEhowto:Cheong}. Hence, we can also think the $\gamma_R$ and
$\gamma_E$ can be set by the desired secrecy rate of the system.

\subsection{2HR-($r,k$) Protocol}

Notice the available transmission protocols have their own
advantages and disadvantages in terms of the transmission efficiency
and energy consumption, and thus are suitable for different network
scenarios. With respect to these protocols as special cases, a
general transmission protocol 2HR-($r,k$) is proposed to control the
balance of load distributed among the relays and works as follows.

\begin{enumerate}

\item
\textbf{\emph{Relay selection region determination}:} The circle
area, with radius $r$ and the center at the middle point between
source $S$ and destination $D$, is determined as relay selection
region.

\item
\textbf{\emph{Channel measurement}:} The source $S$ and destination
$D$ broadcast a pilot signal to allow each relay to measure the
channel from $S$ and $D$ to itself. The relays, which receive the
pilot signal, can accurately calculate $h_{S,R_j}, j=1,2,\cdots,n$
and $h_{D,R_j}, j=1,2,\cdots,n$.

\item
\textbf{\emph{Candidate relay selection}:} The relays with the first
$k$ large $min\left(|h_{S,R_j^r}|^2, |h_{D,R_j^r}|^2\right)$ form
the candidate relay set $\mathfrak{R}$. Here, $R_j^r$ denotes the
$j$-th relay node in the relay selection region.

\item
\textbf{\emph{Relay selection}:} The relay, indexed by $j^\ast$, is
selected randomly from candidate relay set $\mathfrak{R}$. Using the
same method with Step 2, each of the other relays $R_j,
j=1,2,\cdots,n, j \neq j^\ast$ in network exactly knows
$h_{R_j,R_{j^\ast}}$.

\item
\textbf{\emph{Two-Hop transmission}:} The source $S$ transmits the
message to $R_{j^\ast}$, and concurrently, the relay nodes with
indexes in $\mathcal {R}_1 = {\left\{j \neq j^\ast :
|h_{R_j,R_{j^\ast}}|^2 < \tau \right\}}$ transmit noise to generate
interference at eavesdroppers. The relay $R_{j^\ast}$ then transmits
the message to destination $D$, and concurrently, the relay nodes
with indexes in $\mathcal {R}_2 = {\left\{j \neq j^\ast :
|h_{R_j,D}|^2 < \tau \right\}}$ transmit noise to generate
interference at eavesdroppers.

\end{enumerate}

\emph{Remark 1}: The load is completely balanced among the relays in
the candidate relay set $\mathfrak{R}$ whose size is determined by
parameter $r$ and $k$ in the 2HR-($r,k$) protocol. Notice that a too
larger $r$ and $k$ may lead to larger size of the candidate relay
set $\mathfrak{R}$. Thus, the load-balance can be flexibly
controlled by a proper setting of the parameter $r$ and $k$ in terms
of network performance requirements.

\emph{Remark 2}: The parameter $\tau$ involved in the 2HR-($r,k$)
protocol serves as the threshold on path-loss, based on which the
set of noise generating relay nodes can be identified. Notice that a
too large $\tau$ may disable legitimate transmission, while a too
small $\tau$ may not be sufficient for interrupting all
eavesdroppers. Thus, the parameter $\tau$ should be set properly to
ensure both secrecy requirement and reliability requirement.

\emph{Remark 3}: In the case that there is equal path-loss between
all pairs of nodes, i.e., $d_{A,B} = 1$ for all $A \neq B$, the
channel state information is independent of the parameter $r$ in
2HR-($r,k$) protocol. Since the parameter $r$ is no effect on relay
selection, the relay selection region is the whole network area with
$r = \infty$. Therefore, 2HR-($r,k$) protocol is castrated as
2HR-($\infty,k$) in case of equal path-loss between all node pairs.

\subsection{Transmission Outage and Secrecy Outage}

For a Two-hop relay transmission from the source $S$ to destination
$D$, we call transmission outage happens if $D$ can not receive the
transmitted packet. We define the transmission outage probability,
denoted by $P_{out}^{\left(T\right)}$, as the probability that
transmission outage from $S$ to $D$ happens. For a predefined upper
bound $\varepsilon_t$ on $P_{out}^{\left(T\right)}$, we call the
communication between $S$ and $D$ is reliable if
$P_{out}^{\left(T\right)} \leq \varepsilon_t$. Similarly, we define
the transmission outage events $O_{S \rightarrow R_{j^\ast}}^{(T)}$
and $O_{R_{j^\ast} \rightarrow D}^{(T)}$ for the transmissions from
$S$ to the selected relay $R_{j^\ast}$ and from $R_{j^\ast}$ to $D$,
respectively. Due to the link independence assumption, we have

\begin{equation}
\begin{aligned}
&P_{out}^{\left(T\right)} = P\left(O_{S \rightarrow
R_{j^\ast}}^{(T)}\right)+P\left(O_{R_{j^\ast}
\rightarrow D}^{(T)}\right)\\
&\ \ \ \ \ \ \ \ \ -P\left(O_{S \rightarrow R_{j^\ast}}^{(T)}\right)
\cdot P\left(O_{R_{j^\ast} \rightarrow D}^{(T)}\right)
\end{aligned}
\end{equation}

Regarding the secrecy outage, we call secrecy outage happens for a
transmission from $S$ to $D$ if at least one eavesdropper can
recover the transmitted packets during the process of this two-hop
transmission. We define the secrecy outage probability, denoted by
$P_{out}^{\left(S\right)}$, as the probability that secrecy outage
happens during the transmission from $S$ to $D$. For a predefined
upper bound $\varepsilon_s$ on $P_{out}^{\left(S\right)}$, we call
the communication between $S$ and $D$ is secure if
$P_{out}^{\left(S\right)} \leq \varepsilon_s$. Similarly, we define
the secrecy outage events $O_{S \rightarrow R_{j^\ast}}^{(S)}$ and
$O_{R_{j^\ast} \rightarrow D}^{(S)}$ for the transmissions from $S$
to the selected relay $R_{j^\ast}$ and from $R_{j^\ast}$ to $D$,
respectively. Due to the link independence assumption, we have

\begin{equation}
\begin{aligned}
&P_{out}^{\left(S\right)} =P\left(O_{S \rightarrow
R_{j^\ast}}^{(S)}\right)+P\left(O_{R_{j^\ast} \rightarrow
D}^{(S)}\right)\\
&\ \ \ \ \ \ \ \ \ -P\left(O_{S \rightarrow R_{j^\ast}}^{(S)}\right)
\cdot P\left(O_{R_{j^\ast} \rightarrow D}^{(S)}\right)
\end{aligned}
\end{equation}

\section{Equal Path-Loss Between All Node Pairs}

In this section, we analyze 2HR-($r,k$) protocol in the case where
the path-loss is equal between all pairs of nodes in the system. The
\emph{Remark 3} shows 2HR-($r,k$) protocol is castrated as
2HR-($\infty,k$) in case of equal path-loss between all node pairs.
We now analyze that under the 2HR-($\infty,k$) protocol the number
of eavesdroppers one network can tolerate subject to specified
requirements on transmission outage and secrecy outage. The
following two lemmas regarding some basic properties of
$P_{out}^{\left(T\right)}$, $P_{out}^{\left(S\right)}$ and $\tau$
are first presented, which will help us to derive the main result in
Theorem 1.

\emph{Lemma 1}: Consider the network scenario of Fig 1 with equal
path-loss between all pairs of nodes, under the 2HR-($r,k$) protocol
the transmission outage probability $P_{out}^{\left(T\right)}$ and
secrecy outage probability $P_{out}^{\left(S\right)}$ there satisfy
the following conditions.

\begin{equation}
\begin{aligned}
&P_{out}^{\left(T\right)} \leq 2 \left(\frac{1}{k}\sum
\limits_{j=1}^k \bigg[\sum \limits_{i=n-j+1}^n \binom{n}{i}
\left[1-\Psi\right]^{i}\Psi^{n-i}\bigg]\right)\\
&\ \ \ \ \ \ \ \ \ \ -\left(\frac{1}{k}\sum \limits_{j=1}^k
\bigg[\sum \limits_{i=n-j+1}^n \binom{n}{i}
\left[1-\Psi\right]^{i}\Psi^{n-i}\bigg]\right)^2
\end{aligned}
\end{equation}

here $\Psi =
e^{-2\gamma_R{\left(n-1\right)\left(1-e^{-\tau}\right)}\tau}$, and

\begin{equation}
\begin{aligned}
&P_{out}^{\left(S\right)} \leq 2m \cdot \left(\frac{1}{1+\gamma_E}\right)^{\left(n-1\right)\left(1-e^{-\tau}\right)}\\
&\ \ \ \ \ \ \ \ \ -\left[m \cdot
\left(\frac{1}{1+\gamma_E}\right)^{\left(n-1\right)\left(1-e^{-\tau}\right)}\right]^2
\end{aligned}
\end{equation}

The proof of Lemma 1 can be found in the Appendix A.

\emph{Lemma 2}: Consider the network scenario of Fig 1 with equal
path-loss between all pairs of nodes, to ensure
$P_{out}^{\left(T\right)} \leq \varepsilon_t$ and
$P_{out}^{\left(S\right)} \leq \varepsilon_s$ under the 2HR-($r,k$)
protocol, the parameter $\tau$ must satisfy the following condition.

\begin{align*}
\tau \leq \sqrt{\frac{ -\log\left(
\left[\binom{k}{\lfloor\frac{k}{2}\rfloor}\left(1 +
k\sqrt{1-\varepsilon_t}\right)\right]^{\frac{1}{k}} - 1
\right)}{2\gamma_R\left(n-1\right)}}
\end{align*}
and

\begin{align*}
\tau \geq - \log{\left[1 + \frac{\log{\left(\frac{1 - \sqrt{1 -
\varepsilon_s}}{m}\right)}}{\left(n - 1\right)\log{\left(1 +
\gamma_E\right)}}\right]}
\end{align*}

here, $\lfloor \cdot \rfloor$ is the floor function.

\begin{proof}

The parameter $\tau$ should be set properly to satisfy both
reliability and secrecy requirements.

\textbf{$\bullet$ Reliability Guarantee}

To ensure the reliability requirement $P_{out}^{\left(T\right)} \leq
\varepsilon_t$, we know from formula (3) in the Lemma 1, that we
just need

\begin{align*}
&2 \left(\frac{1}{k}\sum \limits_{j=1}^k \bigg[\sum
\limits_{i=n-j+1}^n \binom{n}{i}
\left[1-\Psi\right]^{i}\Psi^{n-i}\bigg]\right)\\
&-\left(\frac{1}{k}\sum \limits_{j=1}^k \bigg[\sum
\limits_{i=n-j+1}^n \binom{n}{i}
\left[1-\Psi\right]^{i}\Psi^{n-i}\bigg]\right)^2\\
&\leq \varepsilon_t
\end{align*}

Thus,

\begin{equation}
\begin{aligned}
&\frac{1}{k}\sum \limits_{j=1}^k \bigg[\sum \limits_{i=n-j+1}^n
\binom{n}{i} \left[1-\Psi\right]^{i}\Psi^{n-i}\bigg] \leq 1-
\sqrt{1-\varepsilon_t}
\end{aligned}
\end{equation}

Notice that

\begin{equation}
\begin{aligned}
&\frac{1}{k}\sum \limits_{j=1}^k \bigg[\sum \limits_{i=n-j+1}^n
\binom{n}{i}\left(1-\Psi\right)^{i}\Psi^{n-i}\bigg]\\
& = \frac{1}{k}\sum \limits_{j=1}^k \bigg[1 - \sum
\limits_{i=0}^{n-j}
\binom{n}{i}\left(1-\Psi\right)^{i}\Psi^{n-i}\bigg]\\
& = \frac{1}{k}\sum \limits_{j=1}^k \bigg[1 - \sum
\limits_{i=0}^{n-j}
\frac{\binom{n}{i}}{\binom{n-j}{i}}\binom{n-j}{i}\left(1-\Psi\right)^{i}\Psi^{n-j-i
}\Psi^j\bigg]
\end{aligned}
\end{equation}

We also notice the $i$ can take from $0$ to $n-j$, then we have

\begin{equation*}
\begin{aligned}
& 1 \leq \frac{\binom{n}{i}}{\binom{n-j}{i}} \leq \frac{n!}{(n-j)!j!}\\
\end{aligned}
\end{equation*}

Substituting into formula (6), we have

\begin{equation}
\begin{aligned}
&\frac{1}{k}\sum \limits_{j=1}^k \bigg[1 - \sum \limits_{i=0}^{n-j}
\frac{\binom{n}{i}}{\binom{n-j}{i}}\binom{n-j}{i}\left(1-\Psi\right)^{i}\Psi^{n-j-i
}\Psi^j\bigg]\\
& \leq \frac{1}{k}\sum \limits_{j=1}^k \bigg[1 - \Psi^j \cdot \sum
\limits_{i=0}^{n-j}\binom{n-j}{i}\left(1-\Psi\right)^{i}\Psi^{n-j-i}\bigg]\\
&= 1 - \frac{1}{k}\sum \limits_{j=1}^k  \Psi^j \\
&= 1 - \frac{1}{k}\bigg[\sum \limits_{j=0}^k \frac{1}{\binom{k}{j}} \binom{k}{j} \Psi^j - 1\bigg]\\
&\leq 1 - \frac{1}{k}\bigg[\frac{1}{\binom{k}{\lfloor\frac{k}{2}\rfloor}} \sum \limits_{j=0}^k  \binom{k}{j} \Psi^j - 1\bigg]\\
&= 1 - \frac{1}{k}\bigg[\frac{1}{\binom{k}{\lfloor\frac{k}{2}\rfloor}} (1+\Psi)^k - 1\bigg]\\
\end{aligned}
\end{equation}

According to formula (5), (6) and (7), in order to ensure the
reliability, we need

\begin{equation*}
1 - \frac{1}{k}\bigg[\frac{1}{\binom{k}{\lfloor\frac{k}{2}\rfloor}}
(1+\Psi)^k - 1\bigg] \leq 1- \sqrt{1-\varepsilon_t}
\end{equation*}

or equally,

\begin{equation*}
\Psi \geq \left[\binom{k}{\lfloor\frac{k}{2}\rfloor}\left(1 +
k\sqrt{1-\varepsilon_t}\right)\right]^{\frac{1}{k}} - 1
\end{equation*}

that is,

\begin{equation*}
e^{-2\gamma_R\left(n-1\right) \cdot \left(1-e^{-\tau}\right)\tau}
\geq \left[\binom{k}{\lfloor\frac{k}{2}\rfloor}\left(1 +
k\sqrt{1-\varepsilon_t}\right)\right]^{\frac{1}{k}} - 1
\end{equation*}

Therefore

\begin{equation*}
\left(1-e^{-\tau}\right)\tau \leq \frac{ -\log\left(
\left[\binom{k}{\lfloor\frac{k}{2}\rfloor}\left(1 +
k\sqrt{1-\varepsilon_t}\right)\right]^{\frac{1}{k}} - 1
\right)}{2\gamma_R\left(n-1\right) }
\end{equation*}

By using Taylor formula, we have

\begin{align*}
\tau \leq \sqrt{\frac{ -\log\left(
\left[\binom{k}{\lfloor\frac{k}{2}\rfloor}\left(1 +
k\sqrt{1-\varepsilon_t}\right)\right]^{\frac{1}{k}} - 1
\right)}{2\gamma_R\left(n-1\right)}}
\end{align*}

\textbf{$\bullet$ Secrecy Guarantee}

To ensure the secrecy requirement $P_{out}^{\left(S\right)} \leq
\varepsilon_s$, we know from Lemma 1 that we just need

\begin{align*}
&2m \cdot \left(\frac{1}{1+\gamma_E}\right)^{\left(n-1\right)\left(1-e^{-\tau}\right)}\\
& -\left[m \cdot
\left(\frac{1}{1+\gamma_E}\right)^{\left(n-1\right)\left(1-e^{-\tau}\right)}\right]^2\\
& \leq \varepsilon_s
\end{align*}

Thus,

\begin{align*}
&m \cdot
\left(\frac{1}{1+\gamma_E}\right)^{\left(n-1\right)\left(1-e^{-\tau}\right)}
\leq 1- \sqrt{1-\varepsilon_s}
\end{align*}

That is,

\begin{align*}
& \tau \geq - \log{\left[1 + \frac{\log{\left(\frac{1 - \sqrt{1 -
\varepsilon_s}}{m}\right)}}{\left(n - 1\right)\log{\left(1 +
\gamma_E\right)}}\right]}
\end{align*}

\end{proof}

Based on the results of Lemma 2, we now can establish the following
theorem regarding the performance of the proposed protocol in case
of equal path-loss between all node pairs.

\textbf{Theorem 1.}  Consider the network scenario of Fig 1 with
equal path-loss between all pairs of nodes. To guarantee
$P_{out}^{\left(T\right)} \leq \varepsilon_t$ and
$P_{out}^{\left(S\right)} \leq \varepsilon_s$ under 2HR-($r,k$)
protocol, the number of eavesdroppers $m$ the network can tolerate
must satisfy the following condition.

\begin{align*}
& m \leq \frac{1 - \sqrt{1 -
\varepsilon_s}}{\left(\frac{1}{1+\gamma_E}\right)^{\sqrt{\frac{-\left(n-1\right)\log\left(
\left[\binom{k}{\lfloor\frac{k}{2}\rfloor}\left(1 +
k\sqrt{1-\varepsilon_t}\right)\right]^{\frac{1}{k}} - 1
\right)}{2\gamma_R}}}}
\end{align*}

\begin{proof}

From Lemma 2, we know that to ensure the reliability requirement, we
have

\begin{equation}
\begin{aligned}
\tau \leq \sqrt{\frac{ -\log\left(
\left[\binom{k}{\lfloor\frac{k}{2}\rfloor}\left(1 +
k\sqrt{1-\varepsilon_t}\right)\right]^{\frac{1}{k}} - 1
\right)}{2\gamma_R\left(n-1\right)}}
\end{aligned}
\end{equation}

and

\begin{equation}
\begin{aligned}
\left(n-1\right)\left(1-e^{-\tau}\right) \leq \frac{ -\log\left(
\left[\binom{k}{\lfloor\frac{k}{2}\rfloor}\left(1 +
k\sqrt{1-\varepsilon_t}\right)\right]^{\frac{1}{k}} - 1
\right)}{2\gamma_R\tau}
\end{aligned}
\end{equation}

To ensure the secrecy requirement, we need

\begin{equation}
\begin{aligned}
& \left(\frac{1}{1+\gamma_E}\right)^{\left(n
-1\right)\left(1-e^{-\tau}\right)} \leq \frac{1 - \sqrt{1 -
\varepsilon_s}}{m}
\end{aligned}
\end{equation}

From formula (9) and (10), we can get

\begin{equation}
\begin{aligned}
& m \leq \frac{1 - \sqrt{1 -
\varepsilon_s}}{\left(\frac{1}{1+\gamma_E}\right)^{\left(n
-1\right)\left(1-e^{-\tau}\right)}}\\
&\ \ \ \ \leq \frac{1 - \sqrt{1 -
\varepsilon_s}}{\left(\frac{1}{1+\gamma_E}\right)^{\frac{
-\log\left( \left[\binom{k}{\lfloor\frac{k}{2}\rfloor}\left(1 +
k\sqrt{1-\varepsilon_t}\right)\right]^{\frac{1}{k}} - 1
\right)}{2\gamma_R\tau}}}
\end{aligned}
\end{equation}

By letting $\tau$ take its maximum value for maximum interference at
eavesdroppers, from formula (8) and (11), we get the following bound

\begin{align*}
& m \leq \frac{1 - \sqrt{1 -
\varepsilon_s}}{\left(\frac{1}{1+\gamma_E}\right)^{\sqrt{\frac{-\left(n-1\right)\log\left(
\left[\binom{k}{\lfloor\frac{k}{2}\rfloor}\left(1 +
k\sqrt{1-\varepsilon_t}\right)\right]^{\frac{1}{k}} - 1
\right)}{2\gamma_R}}}}
\end{align*}

\end{proof}

Based on the above analysis, by simple derivation, we can get the
follow corollary to show our proposal is a general protocol.

\textbf{Corollary 1.} Consider the network scenario of Fig 1 with
equal path-loss between all pairs of nodes, the analysis results of
the proposed protocol is identical to that of protocols with the
optimal relay selection presented in
\cite{IEEEhowto:Goeckel1}\cite{IEEEhowto:Goeckel2} by setting of $k
= 1$ and $r = \infty$, and is identical to that of protocols with
the random relay selection presented in
\cite{IEEEhowto:Shen1}\cite{IEEEhowto:Shen2} by setting of $k = n$
and $r = \infty$.

\emph{Remark 4}: In case of equal path-loss of all pairs of nodes
and the parameter $r = \infty$, we notice that the larger $k$ means
the better load-balance among the relays and the lower transmission
efficiency, and vice versa. In particular, when $k=1$, 2HR-($r,k$)
protocol has the worse load-balance among the relays and the highest
transmission efficiency, and when $k=n$, 2HR-($r,k$) protocol has
the best load-balance among the relays and the lower transmission
efficiency.

\section{General Case To Addressing Path-Loss}

In this section, we consider the more general scenario where the
path-loss between each pair of nodes also depends on the distance
between them. The related theoretic analysis is further provided to
determine the number of eavesdroppers one network can tolerant by
adopting the 2HR-($r,k$) protocol. To address the distance-dependent
path-loss, we consider a coordination system shown in Fig 2, in
which the two-hop relay wireless networks employed in the 2-D plane
of unit area, consisting of the square $\left[-0.5, 0.5\right]
\times \left[-0.5, 0.5\right]$. The source $S$ located at coordinate
$\left(-0.5,0\right)$ wishes to establish two-hop transmission with
destination $D$ located at coordinate $\left(0.5,0\right)$.

\begin{figure}[!t]
\centering
\includegraphics[width=3.7in]{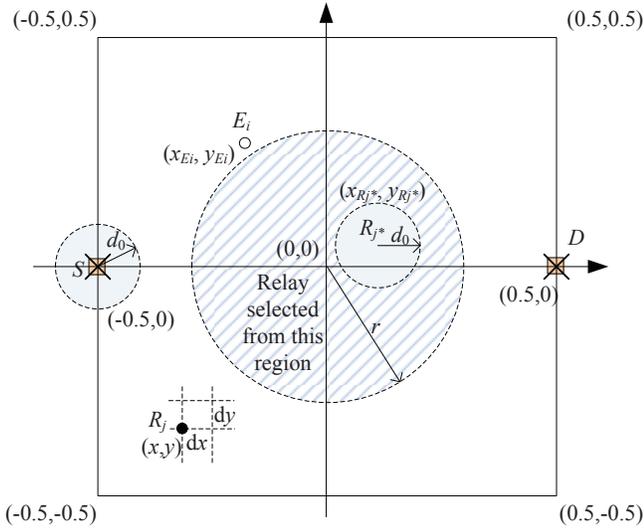}
\DeclareGraphicsExtensions. \caption{Coordinate system for the
scenario where path-loss between pairs of nodes is based on their
relative locations.} \label{Coordinate system}
\end{figure}

To address the near eavesdropper problem and also to simply the
analysis for the 2HR-($r,k$) protocol, we assume that there exits a
constant $d_0>0$ such that any eavesdropper falling within a circle
area with radius $d_0$ and center $S$ or $R_{j^\ast}$ can eavesdrop
the transmitted messages successfully with probability 1, while any
eavesdropper beyond such area can only successfully eavesdropper the
transmitted messages with a probability less than 1. Based on such a
simplification, we can establish the following two lemmas regarding
some basic properties of $P_{out}^{\left(T\right)}$,
$P_{out}^{\left(S\right)}$ and $\tau$ under this protocol.

\emph{Lemma 3}: Consider the network scenario of Fig 2, under the
2HR-($r,k$) protocol the transmission outage probability
$P_{out}^{\left(T\right)}$ and secrecy outage probability
$P_{out}^{\left(S\right)}$ there satisfy the following condition.

\begin{equation}
\begin{aligned}
&P_{out}^{\left(T\right)} \leq 1 - \Upsilon^{\varphi_1+\varphi_2}
\sum \limits_{l=1}^k \binom{n}{l} \left(\pi r^2\right)^l\left(1 -
\pi
r^2\right)^{n-l} \\
&\ \ \ \ \ \ \ \ \ - \frac{\Upsilon^{2\left(\varphi_1 +
\varphi_2\right)}}{k^2} \sum \limits_{l=k+1}^n \binom{n}{l}
\left(\pi r^2\right)^l\left(1 - \pi r^2\right)^{n-l}\\
\end{aligned}
\end{equation}

\begin{equation}
\begin{aligned}
&P_{out}^{\left(S\right)} \leq \\
&2 m \left[\pi {d_0}^2 + \left(\frac{1}{1 + \gamma_E \psi
{d_0}^{\alpha}}\right)^{\left(n-1\right)\left(1-e^{-\tau}\right)}\left(1- \pi {d_0}^2\right)\right]\\
&-\left[m \left(\pi {d_0}^2 + \left(\frac{1}{1 + \gamma_E \psi
{d_0}^{\alpha}}\right)^{\left(n-1\right)\left(1-e^{-\tau}\right)}\left(1-
\pi {d_0}^2\right)\right)\right]^2
\end{aligned}
\end{equation}

here,

\begin{align*}
&\Upsilon = e^{-\frac{\gamma_R {\tau \left(n-1\right)\left(1 -
e^{-\tau}\right)}}{\left(0.5+r\right)^{-\alpha}}}
\end{align*}

\begin{align*}
&\varphi_1 = \int_{-0.5}^{0.5}\int_{-0.5}^{0.5}\frac{1}{\left(x^2 +
y^2\right)^{\frac{\alpha}{2}}}dxdy
\end{align*}

\begin{align*}
&\varphi_2
=\int_{-0.5}^{0.5}\int_{-0.5}^{0.5}\frac{1}{\left[\left(x-0.5\right)^2
+ y^2\right]^{\frac{\alpha}{2}}}dxdy
\end{align*}

\begin{align*}
&\psi =
\int_{-0.5}^{0.5}\int_{-0.5}^{0.5}\frac{1}{\left[\left(x-0.5\right)^2
+ \left(y-0.5\right)^2\right]^{\frac{\alpha}{2}}}dxdy
\end{align*}

The proof of Lemma 3 can be found in the Appendix B.

\emph{Lemma 4}: Consider the network scenario of Fig 2, to ensure
$P_{out}^{\left(T\right)} \leq \varepsilon_t$ and
$P_{out}^{\left(S\right)} \leq \varepsilon_s$ by applying
2HR-($r,k$) protocol, the parameter $\tau$ must satisfy the
following condition.

\begin{align*}
\tau \leq \sqrt{\frac{-\log{\left[\frac{k^2\sqrt{{\nu_1}^2 +
4\left(1-\varepsilon_t\right)\nu_2}-k^2\nu_1}{2\nu_2}\right]}}{\gamma_R\left(n-1\right)\left(\varphi_1+\varphi_2\right)\left(0.5+r\right)^{\alpha}}}
\end{align*}

and

\begin{equation*}
\tau \geq - \log\left[1 + \frac{\log{\left(\frac{\frac{1 - \sqrt{1 -
\varepsilon_s}}{m} - \pi {d_0}^2}{1-\pi
{d_0}^2}\right)}}{\left(n-1\right)\log{\left(1 + \gamma_E \psi
{d_0}^{\alpha}\right)}}\right]
\end{equation*}

here, $\varphi_1$, $\varphi_2$, and $\psi$ are defined in the same
way as that in Lemma 3, and

\begin{align*}
&\nu_1 = k^2\sum \limits_{l=1}^k \binom{n}{l} \left(\pi
r^2\right)^l\left(1 - \pi r^2\right)^{n-l}
\end{align*}

\begin{align*}
&\nu_2 = k^2\sum \limits_{l=k+1}^n \binom{n}{l} \left(\pi
r^2\right)^l\left(1 - \pi r^2\right)^{n-l}
\end{align*}

\begin{proof}

The parameter $\tau$ should be set properly to satisfy both
reliability and secrecy requirements.

\textbf{$\bullet$ Reliability Guarantee}

To ensure the reliability requirement $P_{out}^{\left(T\right)} \leq
\varepsilon_t$, we know from formula (12) in Lemma 3 that we just
need

\begin{equation*}
\begin{aligned}
&1 - \Upsilon^{\varphi_1+\varphi_2} \sum \limits_{l=1}^k
\binom{n}{l} \left(\pi r^2\right)^l\left(1 - \pi
r^2\right)^{n-l} \\
&\ \ \ \ \ \ \ \ \ - \frac{\Upsilon^{2\left(\varphi_1 +
\varphi_2\right)}}{k^2} \sum \limits_{l=k+1}^n \binom{n}{l}
\left(\pi r^2\right)^l\left(1 - \pi r^2\right)^{n-l}\\
&\leq \varepsilon_t
\end{aligned}
\end{equation*}

Thus,

\begin{equation*}
\begin{aligned}
&\Upsilon^{\varphi_1+\varphi_2} \geq \frac{k^2\sqrt{{\nu_1}^2 +
4\left(1-\varepsilon_t\right)\nu_2}-k^2\nu_1}{2\nu_2}
\end{aligned}
\end{equation*}

here

\begin{align*}
&\nu_1 = k^2\sum \limits_{l=1}^k \binom{n}{l} \left(\pi
r^2\right)^l\left(1 - \pi r^2\right)^{n-l}
\end{align*}

\begin{align*}
&\nu_2 = k^2\sum \limits_{l=k+1}^n \binom{n}{l} \left(\pi
r^2\right)^l\left(1 - \pi r^2\right)^{n-l}
\end{align*}

That is,

\begin{equation*}
\begin{aligned}
&e^{-\frac{\gamma_R \tau \left(n-1\right)\left(1 -
e^{-\tau}\right)\left(\varphi_1+\varphi_2\right)}{\left(0.5+r\right)^{-\alpha}}}\\
&\ \ \ \ \ \ \ \ \ \ \geq \frac{k^2\sqrt{{\nu_1}^2 +
4\left(1-\varepsilon_t\right)\nu_2}-k^2\nu_1}{2\nu_2}
\end{aligned}
\end{equation*}

Thus,

\begin{equation*}
\begin{aligned}
& \tau \left(1 - e^{-\tau}\right) \leq
\frac{-\log{\left[\frac{k^2\sqrt{{\nu_1}^2 +
4\left(1-\varepsilon_t\right)\nu_2}-k^2\nu_1}{2\nu_2}\right]}}{\gamma_R\left(n-1\right)\left(\varphi_1+\varphi_2\right)\left(0.5+r\right)^{\alpha}}
\end{aligned}
\end{equation*}

By using Taylor formula, we have

\begin{align*}
\tau \leq \sqrt{\frac{-\log{\left[\frac{k^2\sqrt{{\nu_1}^2 +
4\left(1-\varepsilon_t\right)\nu_2}-k^2\nu_1}{2\nu_2}\right]}}{\gamma_R\left(n-1\right)\left(\varphi_1+\varphi_2\right)\left(0.5+r\right)^{\alpha}}}
\end{align*}

\textbf{$\bullet$ Secrecy Guarantee}

To ensure the secrecy requirement $P_{out}^{\left(S\right)} \leq
\varepsilon_s$, we know from formula (13) in Lemma 3 that we just
need

\begin{equation*}
\begin{aligned}
&2 m \left[\pi {d_0}^2 + \left(\frac{1}{1 + \gamma_E \psi
{d_0}^{\alpha}}\right)^{\left(n-1\right)\left(1-e^{-\tau}\right)}\left(1- \pi {d_0}^2\right)\right]\\
&-\left[m \left(\pi {d_0}^2 + \left(\frac{1}{1 + \gamma_E \psi
{d_0}^{\alpha}}\right)^{\left(n-1\right)\left(1-e^{-\tau}\right)}\left(1-
\pi {d_0}^2\right)\right)\right]^2\\
& \leq \varepsilon_s
\end{aligned}
\end{equation*}

Thus,

\begin{equation*}
\begin{aligned}
& m \cdot \left[\pi {d_0}^2 + \left(\frac{1}{1 + \gamma_E \psi
{d_0}^{\alpha}}\right)^{\left(n-1\right)\left(1-e^{-\tau}\right)} \left(1- \pi {d_0}^2\right) \right]\\
& \ \ \ \ \ \ \ \ \ \ \ \ \ \ \ \ \ \ \ \ \ \ \ \ \ \ \ \ \ \ \ \
\leq 1- \sqrt{1-\varepsilon_s}
\end{aligned}
\end{equation*}

that is,

\begin{equation*}
\tau \geq - \log\left[1 + \frac{\log{\left(\frac{\frac{1 - \sqrt{1 -
\varepsilon_s}}{m} - \pi {d_0}^2}{1-\pi
{d_0}^2}\right)}}{\left(n-1\right)\log{\left(1 + \gamma_E \psi
{d_0}^{\alpha}\right)}}\right]
\end{equation*}

\end{proof}

Based on the results of Lemma 4, we now can establish the following
theorem regarding the performance of 2HR-($r,k$) protocol.

\textbf{Theorem 2.} Consider the network scenario of Fig 2. To
guarantee $P_{out}^{\left(T\right)} \leq \varepsilon_t$ and
$P_{out}^{\left(S\right)} \leq \varepsilon_s$ based on the proposed
2HR-($r,k$) protocol, the number of eavesdroppers $m$ the network
can tolerate must satisfy the following condition.

\begin{equation*}
\begin{aligned}
& m \leq \frac{1- \sqrt{1-\varepsilon_s}}{\pi {d_0}^2 + \left(1- \pi
{d_0}^2\right)\omega}
\end{aligned}
\end{equation*}

here

\begin{align*}
& \omega = \left(\frac{1}{1 + \gamma_E \psi
{d_0}^{\alpha}}\right)^{\sqrt{\frac{-\left(n-1\right)\log{\left[\frac{k^2\sqrt{{\nu_1}^2
+
4\left(1-\varepsilon_t\right)\nu_2}-k^2\nu_1}{2\nu_2}\right]}}{\gamma_R\left(\varphi_1+\varphi_2\right)\left(0.5+r\right)^{\alpha}}}}
\end{align*}

$\varphi_1$, $\varphi_2$, $\nu_1$,$\nu_2$ and $\psi$ are defined in
the same way as that in Lemma 3 and Lemma 4.

\begin{proof}

From Lemma 4, we know that to ensure the reliability requirement, we
have

\begin{equation}
\begin{aligned}
\tau \leq \sqrt{\frac{-\log{\left[\frac{k^2\sqrt{{\nu_1}^2 +
4\left(1-\varepsilon_t\right)\nu_2}-k^2\nu_1}{2\nu_2}\right]}}{\gamma_R\left(n-1\right)\left(\varphi_1+\varphi_2\right)\left(0.5+r\right)^{\alpha}}}
\end{aligned}
\end{equation}

and

\begin{equation}
\begin{aligned}
&  \left(n-1\right)\left(1 - e^{-\tau}\right) \leq
\frac{-\log{\left[\frac{k^2\sqrt{{\nu_1}^2 +
4\left(1-\varepsilon_t\right)\nu_2}-k^2\nu_1}{2\nu_2}\right]}}{\gamma_R\tau\left(\varphi_1+\varphi_2\right)\left(0.5+r\right)^{\alpha}}
\end{aligned}
\end{equation}

To ensure the secrecy requirement, we need

\begin{equation}
\begin{aligned}
& m \cdot \left[\pi {d_0}^2 + \left(\frac{1}{1 + \gamma_E \psi
{d_0}^{\alpha}}\right)^{\left(n-1\right)\left(1-e^{-\tau}\right)} \left(1- \pi {d_0}^2\right) \right]\\
& \ \ \ \ \ \ \ \ \ \ \ \ \ \ \ \ \ \ \ \ \ \ \ \ \ \ \ \ \ \ \ \
\leq 1- \sqrt{1-\varepsilon_s}
\end{aligned}
\end{equation}

From formula (15) and (16), we can get

\begin{equation}
\begin{aligned}
& m \leq \frac{1- \sqrt{1-\varepsilon_s}}{\pi {d_0}^2 +
\left(\frac{1}{1 + \gamma_E \psi
{d_0}^{\alpha}}\right)^{\left(n-1\right)\left(1-e^{-\tau}\right)} \left(1- \pi {d_0}^2\right)}\\
&\leq \frac{1- \sqrt{1-\varepsilon_s}}{\pi {d_0}^2 +
\left(\frac{1}{1 + \gamma_E \psi
{d_0}^{\alpha}}\right)^{\frac{-\log{\left[\frac{k^2\sqrt{{\nu_1}^2 +
4\left(1-\varepsilon_t\right)\nu_2}-k^2\nu_1}{2\nu_2}\right]}}{\gamma_R\tau\left(\varphi_1+\varphi_2\right)\left(0.5+r\right)^{\alpha}}}
\left(1- \pi {d_0}^2\right)}
\end{aligned}
\end{equation}

By letting $\tau$ take its maximum value for maximum interference at
eavesdroppers, from formula (14) and (17), we get the following
bound

\begin{equation*}
\begin{aligned}
& m \leq \frac{1- \sqrt{1-\varepsilon_s}}{\pi {d_0}^2 + \left(1- \pi
{d_0}^2\right)\omega}
\end{aligned}
\end{equation*}

here

\begin{align*}
& \omega = \left(\frac{1}{1 + \gamma_E \psi
{d_0}^{\alpha}}\right)^{\sqrt{\frac{-\left(n-1\right)\log{\left[\frac{k^2\sqrt{{\nu_1}^2
+
4\left(1-\varepsilon_t\right)\nu_2}-k^2\nu_1}{2\nu_2}\right]}}{\gamma_R\left(\varphi_1+\varphi_2\right)\left(0.5+r\right)^{\alpha}}}}
\end{align*}

\end{proof}

\emph{Remark 5}: The parameter $r$ determines the relay selection
region. When parameter $r$ tends to $0$, few system nodes locate in
relay selection region, and the relay selection process tends to
optimal from the view of relay selection region with less
load-balance capacity. With increasing of parameter $r$, the more
relays are in relay selection region, which can ensure better
load-balance.

\emph{Remark 6}: The SINR at the receiver depends on channel state
information and the distance between the transmitter and receiver.
The \emph{Remark 4} and \emph{Remark 5} show that the parameter $r$
and $k$ in 2HR-($r,k$) protocol can be flexibly set to control the
tradeoff the load-balance and the transmission efficiency in terms
of channel state information and the distance between the
transmitter and receiver respectively.

\emph{Remark 7}: In order to get the better load-balance, set a
larger $r$ and $k$ which will result in a lower transmission
efficiency. The Theorem 1 and Theorem 3 show that the number of
eavesdroppers one network can tolerant is decreasing as the
increasing $r$ and $k$.

\emph{Remark 8}: In the initial stage of the network operation, the
parameter $r$ and $k$ can be set small values to ensure the high
efficiency, since all relays are energetic which load-balance among
the relays is not first considered. With the passage of time of the
network operation, the parameter $r$ and $k$ can be gradually set
higher values for better load-balance among the relays to extend the
network lifetime.

Based on the above analysis, by simple derivation, we can get the
follow corollary to show our proposal is a general protocol.

\textbf{Corollary 2.} Consider the network scenario of Fig 2, the
analysis results of the proposed protocol with $r \rightarrow
\infty$ and $k=n$ is identical to that of Protocol 3 with $a=0$ and
$b=0$ (the parameters $a$ and $b$ determine the relay selection
region) proposed in \cite{IEEEhowto:Shen2}, and the analysis results
of the proposed protocol with $r \rightarrow 0$ and $k=n$ is
identical to that of Protocol 3 with $a \rightarrow 0.5$ and $b
\rightarrow 0.5$ proposed in \cite{IEEEhowto:Shen2}.

\emph{Remark 9}: The protocol proposed in \cite{IEEEhowto:Shen2}
have the ability to control load-balance among the relays by only
control on the relay selection region. Whereas, 2HR-($r,k$) protocol
can realize load-balance by control on both relay selection set and
relay selection region.

\section{Related Works}

A lot of research works have been dedicated to load-balance
transmission scheme for balanced energy consumption among system
nodes to prolong the network lifetime in wireless networks. A few
dynamic load balancing strategies and schemes were proposed in
\cite{IEEEhowto:Dalalah}\cite{IEEEhowto:Hac} for distributed
systems. For wireless mesh network, a multi-hop transmission scheme
is proposed in \cite{IEEEhowto:Gumel}, in which information relay is
selected based on the current load of the relay nodes. For wireless
access networks, a distributed routing algorithm that performs
dynamic load-balance by constructs a load-balanced backbone tree
\cite{IEEEhowto:Hsiao}. J. Gao et al. extended the shortest path
routing to support load-balance \cite{IEEEhowto:Gao}. In particular,
for energy constrained wireless sensor networks, load-balance is
significant important, and a lot of transmission schemes were
proposed for load-balance among relays and prolonging the network
lifetime
\cite{IEEEhowto:Trajcevski}\cite{IEEEhowto:Wajgi1}\cite{IEEEhowto:Wajgi2}.
Lifetime optimization and security of multi-hop wireless networks
was further considered and the secure transmission scheme with
load-balance is proposed in
\cite{IEEEhowto:Zhang}\cite{IEEEhowto:Ozdemir}.

Recently, attention is turning to achieve physical layer secrecy and
secure transmission scheme via cooperative relays is considered in
large wireless networks. Some transmission protocols are proposed to
select the optimal relay in terms of the maximum secrecy capacity or
minimum transmit power. In case that eavesdropper channels or
locations is known, node cooperation is used to improve the
performance of secure wireless communications and a few cooperative
transmission protocols were proposed to jam eavesdroppers
\cite{IEEEhowto:Dong2}\cite{IEEEhowto:Dong3}. In case that
eavesdropper channels or locations is unknown, D. Goeckel et al.
proposed a transmission protocol based on optimal relay selection
\cite{IEEEhowto:Goeckel1}\cite{IEEEhowto:Goeckel2}. For both
one-dimensional and two-dimensional networks, a secure transmission
protocol is proposed in \cite{IEEEhowto:Capar1}. Z. Ding et al.
considered the opportunistic use of relays and proposed two secrecy
transmission protocols \cite{IEEEhowto:Ding}. The "two-way secrecy
scheme" was studied in \cite{IEEEhowto:Leow} \cite{IEEEhowto:Capar3}
and M. Dehghan et al. explored the energy efficiency of cooperative
jamming scheme \cite{IEEEhowto:Dehghan}. A. Sheikholeslami et al.
proposed a protocol, where the signal of a given transmitter is
protected by the aggregate interference produced by the other
transmitters \cite{IEEEhowto:Sheikholeslami}. A secure transmission
protocol are presented in case where the eavesdroppers collude
\cite{IEEEhowto:Vasudevan2}. J. Li et al. proposed two secure
transmission protocols to confound the eavesdroppers
\cite{IEEEhowto:Li1}. The above works mainly focus on the maximum
the secrecy capacity, in which the system nodes with best link
condition is always selected as information relay. Such, these
protocols have less load-balance capacity. In order to address this
problem, Y. Shen et al. further proposed a protocol with random
relay selection in \cite{IEEEhowto:Shen1}\cite{IEEEhowto:Shen2}.
This protocol can provide good load-balance capacity and balanced
energy consumption among the relays, whereas it has low transmission
efficiency.

\section{Conclusion}

This paper proposed a general 2HR-($r,k$) protocol to ensure secure
and reliable information transmission through multiple cooperative
system nodes for two-hop relay wireless networks without the
knowledge of eavesdropper channels and locations. We proved that the
2HR-($r,k$) protocol has the capability of flexible control over the
tradeoff between the load-balance capacity and the transmission
efficiency by a proper setting of the radius $r$ of relay selection
region and the size $k$ of candidate relay set. Such, in general it
is possible for us to set proper value of parameters according to
network scenario to support various applications. The results in
this paper indicate that the parameters $r$ and $k$ of the
2HR-($r,k$) protocol do also affect the number of eavesdroppers one
networks can tolerant under the premise of specified secure and
reliable requirements.

\appendices

\section{Proof of Lemma 1}

\begin{proof}

Based on the definition of transmission outage probability, we have

\begin{align*}
& P\left(O_{S
\rightarrow R_{j^\ast}}^{(T)}\right)\\
& \ \ \ \ \ = P\left(C_{S,R_{j^\ast}} \leq
\gamma_R\right)\\
& \ \ \ \ \ = P\left(\frac{E_s \cdot |h_{S,R_{j^\ast}}|^2}{\sum_{R_j
\in \mathcal {R}_1}E_s \cdot
|h_{R_j,R_{j^\ast}}|^2 + N_0/2} \leq \gamma_R\right)\\
& \ \ \ \ \ \doteq P\left(\frac{|h_{S,R_{j^\ast}}|^2}{\sum_{R_j \in
\mathcal
{R}_1}|h_{R_j,R_{j^\ast}}|^2} \leq \gamma_R\right)\\
& \ \ \ \ \ \leq P\left(\frac{H}{{|\mathcal {R}_1|}\tau} \leq
\gamma_R\right)\\
& \ \ \ \ \ = P\left(H \leq \gamma_R{|\mathcal {R}_1|}\tau\right)\\
\end{align*}

Here, $H = min \left(|h_{S,R_{j^\ast}}|^2,
|h_{D,R_{j^\ast}}|^2\right)$. Compared to the noise generated by
multiple system nodes, the environment noise is negligible and thus
is omitted here to simply the analysis. Notice that $\mathcal {R}_1
= {\left\{j \neq j^\ast : |h_{R_j,R_{j^\ast}}|^2 < \tau \right\}}$.

Employing Appendix C, we should have

\begin{align*}
& P\left(O_{S \rightarrow R_{j^\ast}}^{(T)}\right) \leq F_{H}\left(\gamma_R{|\mathcal {R}_1|}\tau\right)\\
&\ \ \ \ \ \ \ \ \ \ \ \ \ \ \ \ = \frac{1}{k}\sum \limits_{j=1}^k
\bigg[\sum
\limits_{i=n-j+1}^n \binom{n}{i}\cdot \\
&\ \ \ \ \ \ \ \ \ \ \ \ \ \ \ \ \ \ \
\left[1-e^{-2\gamma_R{|\mathcal
{R}_1|}\tau}\right]^{i}\left[e^{-2\gamma_R{|\mathcal
{R}_1|}\tau}\right]^{n-i}\bigg]\\
\end{align*}

Since there are $n - 1$ other relays except $R_{j^\ast}$, the
expected number of noise-generation nodes is given by $|\mathcal
{R}_1| =\left(n-1\right) \cdot P\left(|h_{R_j,R_{j^\ast}}|^2 <
\tau\right) = \left(n-1\right)\left(1-e^{-\tau}\right)$. Then we
have

\begin{align*}
& P\left(O_{S \rightarrow R_{j^\ast}}^{(T)}\right) \leq
\frac{1}{k}\sum \limits_{j=1}^k \bigg[\sum
\limits_{i=n-j+1}^n \binom{n}{i}\cdot \\
& \left[1-e^{-2\gamma_R{\left(n-1\right)
\left(1-e^{-\tau}\right)}\tau}\right]^{i}\left[e^{-2\gamma_R{\left(n-1\right)\left(1-e^{-\tau}\right)}\tau}\right]^{n-i}\bigg]\\
\end{align*}

For convenience of the description, let $\Psi =
e^{-2\gamma_R{\left(n-1\right) \cdot
\left(1-e^{-\tau}\right)}\tau}$, and we have

\begin{equation}
\begin{aligned}
& P\left(O_{S \rightarrow R_{j^\ast}}^{(T)}\right) \leq
\frac{1}{k}\sum \limits_{j=1}^k \bigg[\sum
\limits_{i=n-j+1}^n \binom{n}{i} \left[1-\Psi\right]^{i}\Psi^{n-i}\bigg]\\
\end{aligned}
\end{equation}

Employing the same method, we can get

\begin{equation}
\begin{aligned}
& P\left(O_{R_{j^\ast} \rightarrow D}^{(T)}\right) \leq
\frac{1}{k}\sum \limits_{j=1}^k \bigg[\sum
\limits_{i=n-j+1}^n \binom{n}{i} \left[1-\Psi\right]^{i}\Psi^{n-i}\bigg]\\
\end{aligned}
\end{equation}

Substituting formula (18) and (19) into formula (1), we have

\begin{align*}
&P_{out}^{\left(T\right)} \leq 2 \left(\frac{1}{k}\sum
\limits_{j=1}^k \bigg[\sum \limits_{i=n-j+1}^n \binom{n}{i}
\left[1-\Psi\right]^{i}\Psi^{n-i}\bigg]\right)\\
&\ \ \ \ \ \ \ \ \ \ -\left(\frac{1}{k}\sum \limits_{j=1}^k
\bigg[\sum \limits_{i=n-j+1}^n \binom{n}{i}
\left[1-\Psi\right]^{i}\Psi^{n-i}\bigg]\right)^2
\end{align*}

According to the definition of secrecy outage probability, we know
that

\begin{align*}
&P\left(O_{S \rightarrow R_{j^\ast}}^{(S)}\right) =
P\left(\bigcup_{i=1}^{m}\left\{C_{S,E_i} \geq
\gamma_E\right\}\right)
\end{align*}

Thus, we have

\begin{equation}
\begin{aligned}
&P\left(O_{S \rightarrow R_{j^\ast}}^{(S)}\right) \leq
\sum_{i=1}^{m}P\left(C_{S,E_i} \geq \gamma_E\right)
\end{aligned}
\end{equation}

Based on the Markov inequality,

\begin{align*}
& P\left(C_{S,E_i} \geq \gamma_E\right)\\
& \ \ \ \ \ \leq P\left(\frac{E_s \cdot |h_{S,E_i}|^2}{\sum_{R_j \in
\mathcal {R}_1}E_s \cdot |h_{R_j,E_i}|^2} \geq
\gamma_E\right)\\
& \ \ \ \ \ = E_{\left\{h_{R_j,E_i}, j=0,1,\cdots,n+mp,j \neq
j^{\ast}\right\},\mathcal {R}_1}\\
& \ \ \ \ \ \ \ \ \ \left[P\left(|h_{S,E_i}|^2 > \gamma_E
\cdot \sum_{R_j \in \mathcal {R}_1}|h_{R_j,E_i}|^2\right)\right]\\
& \ \ \ \ \ \leq E_{\mathcal {R}_1}\left[\prod_{R_j \in \mathcal
{R}_1}
E_{h_{R_j,E_i}}\left[e^{-\gamma_E|h_{R_j,E_i}|^2}\right]\right]\\
& \ \ \ \ \ = E_{\mathcal
{R}_1}\left[\left(\frac{1}{1+\gamma_E}\right)^{|\mathcal
{R}_1|}\right]
\end{align*}

Substituting into formula (20), we have

\begin{equation}
\begin{aligned}
&P\left(O_{S \rightarrow R_{j^\ast}}^{(S)}\right) \leq
\sum_{i=1}^{m}\left(\frac{1}{1+\gamma_E}\right)^{|\mathcal {R}_1|} =
m \cdot \left(\frac{1}{1+\gamma_E}\right)^{|\mathcal {R}_1|}
\end{aligned}
\end{equation}

employing the same method, we can get

\begin{equation}
\begin{aligned}
&P\left(O_{R_{j^\ast} \rightarrow D}^{(S)}\right) \leq m \cdot
\left(\frac{1}{1+\gamma_E}\right)^{|\mathcal {R}_2|}
\end{aligned}
\end{equation}

Since the expected number of noise-generation nodes is given by
$|\mathcal {R}_1| =|\mathcal {R}_2| =
\left(n-1\right)\left(1-e^{-\tau}\right)$, thus, substituting
formula (21) and (22) into formula (2), we can get

\begin{align*}
&P_{out}^{\left(S\right)} \leq 2m \cdot
\left(\frac{1}{1+\gamma_E}\right)^{\left(n-1\right)\left(1-e^{-\tau}\right)}\\
&\ \ \ \ \ \ \ \ \ -\left[m \cdot
\left(\frac{1}{1+\gamma_E}\right)^{\left(n-1\right)\left(1-e^{-\tau}\right)}\right]^2
\end{align*}

\end{proof}

\section{Proof of Lemma 3}

\begin{proof}

Notice that two ways leading to transmission outage are: 1) there
are no candidate relays in the relay selection region; 2) the SINR
at the selected relay or the destination is less than $\gamma_R$. We
also notice that if the number of the eligible relays in candidate
relay region less than or equal to $k$, the relay will be random
selected from candidate relay set $\mathfrak{R}$.

Let $A_l$, $l=0, 1,\cdots, n$, be the event that there are just $l$
system nodes in the relay selection region. We have

\begin{equation}
\begin{aligned}
&P_{out}^{\left(T\right)} = \sum \limits_{l=0}^n
P_{out|A_l}^{\left(T\right)}\cdot P(A_l)
\end{aligned}
\end{equation}

Since the relay is uniformly distributed, the number of relays in
candidate relay region is a binomial distribution $\left(n, \pi
r^2\right)$. We have

\begin{equation}
\begin{aligned}
&P(A_l) = \binom{n}{l} \left(\pi r^2\right)^l\left(1 - \pi
r^2\right)^{n-l}
\end{aligned}
\end{equation}

$P_{out|A_l}^{\left(T\right)}$ is discussed from the following three
aspects.

\textbf{1) $l=0$}

In this case, there are no relays in the relay selection region,
then, we have

\begin{equation}
\begin{aligned}
& P_{out|A_l}^{\left(T\right)} = 1
\end{aligned}
\end{equation}

\textbf{2) $1 \leq l \leq k$}

Since the number of candidate relay nodes is less than or equal to
$k$. The relay selection process is to select relay randomly in the
candidate relay set $\mathfrak{R}$ which consists of these $l$
relays located in the relay selection region.

Notice $P_{out|A_l}^{\left(T\right)}$ is determined as

\begin{equation}
\begin{aligned}
&P_{out|A_l}^{\left(T\right)} = P\left(O_{S \rightarrow
R_{j^\ast}}^{(T)}\bigg|A_l\right)+P\left(O_{R_{j^\ast} \rightarrow
D}^{(T)}\bigg|A_l\right)\\
&\ \ \ \ \ \ \ \ \ \ \ \ -P\left(O_{S \rightarrow
R_{j^\ast}}^{(T)}\bigg|A_l\right) \cdot P\left(O_{R_{j^\ast}
\rightarrow D}^{(T)}\bigg|A_l\right)
\end{aligned}
\end{equation}

Based on the definition of transmission outage probability, we have

\begin{align*}
& P\left(O_{S
\rightarrow R_{j^\ast}}^{(T)}\bigg|A_l\right)\\
& \ \ \ \ \ = P\left(C_{S,R_{j^\ast}} \leq
\gamma_R\bigg|A_l\right)\\
& \ \ \ \ \ = P\left(\frac{E_s \cdot
\frac{|h_{S,R_{j^\ast}}|^2}{d_{S,R_{j^\ast}}^{\alpha}}}{\sum_{R_j
\in \mathcal {R}_1}E_s \cdot
\frac{|h_{R_j,R_{j^\ast}}|^2}{d_{R_j,R_{j^\ast}}^{\alpha}} + \frac{N_0}{2}} \leq \gamma_R\bigg|A_l\right)\\
& \ \ \ \ \ \doteq P\left(\frac{
\frac{|h_{S,R_{j^\ast}}|^2}{d_{S,R_{j^\ast}}^{\alpha}}}{\sum_{R_j
\in \mathcal {R}_1}
\frac{|h_{R_j,R_{j^\ast}}|^2}{d_{R_j,R_{j^\ast}}^{\alpha}}} \leq \gamma_R\bigg|A_l\right)\\
\end{align*}

Compared to the noise generated by multiple system nodes, the
environment noise is negligible and thus is omitted here to simply
the analysis. Notice that $\mathcal {R}_1 = {\left\{j \neq j^\ast :
|h_{R_j,R_{j^\ast}}|^2 < \tau \right\}}$, then

\begin{align*}
& P\left(O_{S \rightarrow R_{j^\ast}}^{(T)}\bigg|A_l\right) \leq
P\left(\frac{|h_{S,R_{j^\ast}}|^2 d_{S,R_{j^\ast}}^{-\alpha}} {
\sum_{R_j \in \mathcal {R}_1} \tau d_{R_j,R_{j^\ast}}^{-\alpha}}
\leq \gamma_R\bigg|A_l\right)\\
\end{align*}

Without loss of generality, Let $\left(x, y\right)$ be the
coordinate of $R_j$, shown in Fig 2. The number of noise generation
nodes in square $\left[x, x+dx\right] \times \left[y, y+dy\right]$
is $\left(n-1\right)\left(1 - e^{-\tau}\right)dxdy$. Then, we have

\begin{align*}
& \sum_{R_j \in \mathcal {R}_1}
\frac{\tau}{d_{R_j,R_{j^\ast}}^{\alpha}} \\
&\ \ \ \ \ \ \ \ = \int_0^1\int_0^1\frac{\tau
\left(n-1\right)\left(1
- e^{-\tau}\right)}{\left[\left(x - x_{R_{j^{\ast}}}\right)^2 + \left(y - y_{R_{j^{\ast}}}\right)^2 \right]^{\frac{\alpha}{2}}}dxdy\\
\end{align*}

where $\left(x_{R_{j^{\ast}}}, y_{R_{j^{\ast}}}\right)$ is the
coordinate of the selected relay $R_{j^{\ast}}$ which locates in the
relay selection region. Because the relays are uniformly
distributed, it is the worst case that the selected relay
$R_{j^{\ast}}$ is located on the point $\left(0, 0\right)$, where
the interference at $R_{j^{\ast}}$ from the noise generation nodes
is largest, and the best case with the selected relay $R_{j^{\ast}}$
located in the edge of the circular relay selection region, where
the interference at $R_{j^{\ast}}$ from the noise generation nodes
is lowest. Then, we consider the worst case and have

$$P\left(O_{S \rightarrow R_{j^\ast}}^{(T)}\bigg|A_l\right) \leq
P\left(\frac{|h_{S,R_{j^\ast}}|^2 d_{S,R_{j^\ast}}^{-\alpha}} { \tau
\left(n-1\right)\left(1 - e^{-\tau}\right) \varphi_1 } \leq
\gamma_R\bigg|A_l\right)$$

here, $$\varphi_1 =
\int_{-0.5}^{0.5}\int_{-0.5}^{0.5}\frac{1}{\left(x^2 +
y^2\right)^{\frac{\alpha}{2}}}dxdy$$

Due to $0.5-r \leq d_{S,R_{j^\ast}} \leq 0.5 + r$, then,

\begin{align*}
& P\left(O_{S \rightarrow R_{j^\ast}}^{(T)}\bigg|A_l\right) \\
&\ \ \ \leq P\left(\frac{|h_{S,R_{j^\ast}}|^2
\left(0.5+r\right)^{-\alpha}} {\tau
\left(n-1\right)\left(1 - e^{-\tau}\right) \varphi_1 } \leq \gamma_R\bigg|A_l\right)\\
&\ \ \ = P\left(|h_{S,R_{j^\ast}}|^2 \leq \frac{\gamma_R {\tau
\left(n-1\right)\left(1 - e^{-\tau}\right)
\varphi_1}}{\left(0.5+r\right)^{-\alpha}}\bigg|A_l\right)\\
&\ \ \ = 1-e^{-\frac{\gamma_R {\tau \left(n-1\right)\left(1 -
e^{-\tau}\right) \varphi_1}}{\left(0.5+r\right)^{-\alpha}}}
\end{align*}

For convenience of description, let $\Upsilon = e^{-\frac{\gamma_R
{\tau \left(n-1\right)\left(1 -
e^{-\tau}\right)}}{\left(0.5+r\right)^{-\alpha}}}$, we have

\begin{equation}
\begin{aligned}
& P\left(O_{S \rightarrow R_{j^\ast}}^{(T)}\bigg|A_l\right) \leq
1-\Upsilon^{\varphi_1}
\end{aligned}
\end{equation}

Employing the same method, we can get

\begin{equation}
\begin{aligned}
& P\left(O_{R_{j^\ast} \rightarrow D}^{(T)}\bigg|A_l\right) \leq 1-
\Upsilon^{\varphi_2}
\end{aligned}
\end{equation}

here,

$$\varphi_2 =\int_{-0.5}^{0.5}\int_{-0.5}^{0.5}\frac{1}{\left[\left(x-0.5\right)^2 +
y^2\right]^{\frac{\alpha}{2}}}dxdy$$

Substituting formula (27) and (28) into formula (26), we have

\begin{equation}
\begin{aligned}
&P_{out|A_l}^{\left(T\right)} \leq \left[1-\Upsilon^{\varphi_1}\right]+\left[1-\Upsilon^{\varphi_2}\right]-\left[1-\Upsilon^{\varphi_1}\right]\left[1-\Upsilon^{\varphi_2}\right]\\
&\ \ \ \ \ \ \ \ \ \ = 1- \Upsilon^{\varphi_1+\varphi_2}
\end{aligned}
\end{equation}

\textbf{3) $k < l \leq n$}

In this case, the relay selection process is to select relay
randomly in the candidate relay set $\mathfrak{R}$ which consists of
the relays with the first $k$ large $min\left(|h_{S,R_j}|^2,
|h_{D,R_j}|^2\right)$ in the relay selection region.

Notice $P_{out|A_l}^{\left(T\right)}$ is determined as

\begin{equation}
\begin{aligned}
&P_{out|A_l}^{\left(T\right)} = P\left(O_{S \rightarrow
R_{j^\ast}}^{(T)}\bigg|A_l\right)+P\left(O_{R_{j^\ast} \rightarrow
D}^{(T)}\bigg|A_l\right)\\
&\ \ \ \ \ \ \ \ \ \ \ \ -P\left(O_{S \rightarrow
R_{j^\ast}}^{(T)}\bigg|A_l\right) \cdot P\left(O_{R_{j^\ast}
\rightarrow D}^{(T)}\bigg|A_l\right)
\end{aligned}
\end{equation}

Let the random variable $H = min\left(|h_{S,R_{j^\ast}}|^2,
|h_{D,R_{j^\ast}}|^2\right)$ and from Appendix C, the distribution
function of $H$ is

\begin{equation}
F_{H}\left(x\right)=\begin{cases} \frac{1}{k}\sum \limits_{j=1}^k
\bigg[ \sum \limits_{i=l-j+1}^l
\binom{l}{i}\\
\ \ \ \ \ \ \
\left[1-e^{-2x}\right]^{i}\left[e^{-2x}\right]^{l-i}\bigg] \ \ \ &
\text{$x > 0$}\\
0 \ \ \ &
\text{$x \leq 0$}\\
\end{cases}
\end{equation}

Based on the definition of transmission outage probability,
employing the similar method above, we have

\begin{equation*}
\begin{aligned}
& P\left(O_{S \rightarrow R_{j^\ast}}^{(T)}\bigg|A_l\right) \\
&\ \ \ \leq P\left(\frac{|h_{S,R_{j^\ast}}|^2
\left(0.5+r\right)^{-\alpha}} {\tau
\left(n-1\right)\left(1 - e^{-\tau}\right) \varphi_1 } \leq \gamma_R\bigg|A_l\right)\\
&\ \ \ \leq P\left(H \leq \frac{\gamma_R {\tau
\left(n-1\right)\left(1 - e^{-\tau}\right)
\varphi_1}}{\left(0.5+r\right)^{-\alpha}}\bigg|A_l\right)\\
\end{aligned}
\end{equation*}

From formula (31), we can get

\begin{equation}
\begin{aligned}
& P\left(O_{S \rightarrow R_{j^\ast}}^{(T)}\bigg|A_l\right) \\
&\ \ \ \leq \frac{1}{k}\sum \limits_{j=1}^k \bigg[\sum
\limits_{i=l-j+1}^l \binom{l}{i} \left(1-\Upsilon^{2\varphi_1}
\right)^{i}\cdot \left(\Upsilon^{2\varphi_1}\right)^{l-i}\bigg]\\
&\ \ \ = \frac{1}{k}\sum \limits_{j=1}^k \bigg[1 - \sum
\limits_{i=0}^{l-j}
\frac{\binom{l}{i}}{\binom{l-j}{i}}\binom{l-j}{i}
\left(1-\Upsilon^{2\varphi_1}
\right)^{i}\cdot \\
&\ \ \ \ \ \ \ \ \ \ \ \ \ \ \ \ \ \ \ \ \ \ \ \ \ \ \ \ \ \ \ \ \ \left(\Upsilon^{2\varphi_1}\right)^{l-j-i}\left(\Upsilon^{2\varphi_1}\right)^j\bigg]\\
&\ \ \ \leq \frac{1}{k}\sum \limits_{j=1}^k \bigg[1 - \left(\Upsilon^{2\varphi_1}\right)^j\bigg]\\
&\ \ \ = 1- \frac{1}{k}\sum \limits_{j=1}^k \left(\Upsilon^{2\varphi_1}\right)^j\\
&\ \ \ = 1- \frac{\Upsilon^{2\varphi_1}\left(1 - \Upsilon^{2k\varphi_1}\right)}{k\left(1-\Upsilon^{2\varphi_1}\right)} \\
\end{aligned}
\end{equation}

Employing the same method, we can get

\begin{equation}
\begin{aligned}
& P\left(O_{R_{j^\ast} \rightarrow D}^{(T)}\bigg|A_l\right) \leq 1-
\frac{\Upsilon^{2\varphi_2}\left(1 -
\Upsilon^{2k\varphi_2}\right)}{k\left(1-\Upsilon^{2\varphi_2}\right)}
\end{aligned}
\end{equation}

Substituting formula (32) and (33) into formula (30), we have

\begin{equation}
\begin{aligned}
& P_{out|A_l}^{\left(T\right)} \leq 1-
\frac{\Upsilon^{2\left(\varphi_1 + \varphi_2\right)}\left(1 -
\Upsilon^{2k\varphi_1}\right)\left(1 -
\Upsilon^{2k\varphi_2}\right)}{k^2\left(1-\Upsilon^{2\varphi_1}\right)\left(1-\Upsilon^{2\varphi_2}\right)}
\end{aligned}
\end{equation}

Substituting formula (24), (25), (29) and (34) into formula (23), we
have

\begin{equation*}
\begin{aligned}
&P_{out}^{\left(T\right)}  = \sum \limits_{l=0}^n
P_{out|A_l}^{\left(T\right)}\cdot P(A_l)\\
&\ \ \ \ \ \ = P_{out|A_0}^{\left(T\right)}\cdot P(A_0) + \sum
\limits_{l=1}^k P_{out|A_l}^{\left(T\right)}\cdot P(A_l)  \\
&\ \ \ \ \ \ \ \ \ + \sum \limits_{l=k+1}^n
P_{out|A_l}^{\left(T\right)}\cdot P(A_l)\\
&\ \ \ \ \ \ \leq 1 \cdot \left(1 - \pi r^2\right)^n \\
& \ \ \ \ \ \ \ \ \ + \left(1-
\Upsilon^{\varphi_1+\varphi_2}\right)\sum \limits_{l=1}^k
\binom{n}{l} \left(\pi r^2\right)^l\left(1 - \pi
r^2\right)^{n-l} \\
&\ \ \ \ \ \ \ \ \ +\left[1- \frac{\Upsilon^{2\left(\varphi_1 +
\varphi_2\right)}\left(1 - \Upsilon^{2k\varphi_1}\right)\left(1 -
\Upsilon^{2k\varphi_2}\right)}{k^2\left(1-\Upsilon^{2\varphi_1}\right)\left(1-\Upsilon^{2\varphi_2}\right)}\right]
\cdot\\
&\ \ \ \ \ \ \ \ \ \ \ \ \sum \limits_{l=k+1}^n \binom{n}{l}
\left(\pi r^2\right)^l\left(1 - \pi r^2\right)^{n-l}\\
\end{aligned}
\end{equation*}

\begin{equation*}
\begin{aligned}
&\ \ \ \ \ \ \leq 1 - \Upsilon^{\varphi_1+\varphi_2} \sum
\limits_{l=1}^k \binom{n}{l} \left(\pi r^2\right)^l\left(1 - \pi
r^2\right)^{n-l} \\
&\ \ \ \ \ \ \ \ \ - \frac{\Upsilon^{2\left(\varphi_1 +
\varphi_2\right)}\left(1 - \Upsilon^{2k\varphi_1}\right)\left(1 -
\Upsilon^{2k\varphi_2}\right)}{k^2\left(1-\Upsilon^{2\varphi_1}\right)\left(1-\Upsilon^{2\varphi_2}\right)}
\cdot\\
&\ \ \ \ \ \ \ \ \ \ \ \ \sum \limits_{l=k+1}^n \binom{n}{l}
\left(\pi r^2\right)^l\left(1 - \pi r^2\right)^{n-l}\\
&\ \ \ \ \ \ \leq 1 - \Upsilon^{\varphi_1+\varphi_2} \sum
\limits_{l=1}^k \binom{n}{l} \left(\pi r^2\right)^l\left(1 - \pi
r^2\right)^{n-l} \\
&\ \ \ \ \ \ \ \ \ - \frac{\Upsilon^{2\left(\varphi_1 +
\varphi_2\right)}}{k^2} \sum \limits_{l=k+1}^n \binom{n}{l}
\left(\pi r^2\right)^l\left(1 - \pi r^2\right)^{n-l}\\
\end{aligned}
\end{equation*}

According to the definition of secrecy outage probability, we know
that

\begin{align*}
&P\left(O_{S \rightarrow R_{j^\ast}}^{(S)}\right) =
P\left(\bigcup_{i=1}^{m}\left\{C_{S,E_i} \geq
\gamma_E\right\}\right)
\end{align*}

Thus, we have

\begin{equation}
\begin{aligned}
&P\left(O_{S \rightarrow R_{j^\ast}}^{(S)}\right) \leq
\sum_{i=1}^{m}P\left(C_{S,E_i} \geq \gamma_E\right)
\end{aligned}
\end{equation}

Based on the definition of $d_0$, we denote by $G_1^{(i)}$ the event
that the distance between $E_i$ and the source is less than $d_0$,
and denote by $G_2^{(i)}$ the event that distance between $E_i$ and
the source is lager than or equal to $d_0$. We have

\begin{equation*}
\begin{aligned}
&P\left(C_{S,E_i} \geq \gamma_E\right) \\
&\ \ \ \ =
P\left(C_{S,E_i} \geq \gamma_E\bigg|G_1^{(i)}\right)P\left(G_1^{(i)}\right)\\
&\ \ \ \ \ \ \ \ +
P\left(C_{S,E_i} \geq \gamma_E\bigg|G_2^{(i)}\right)P\left(G_2^{(i)}\right)\\
&\ \ \ \ \leq 1 \cdot \frac{1}{2} \pi {d_0}^2 \\
&\ \ \ \ \ \ \ \ +
P\left(\frac{\frac{|h_{S,E_i}|^2}{d_{S,E_i}^\alpha}}{\sum\limits_{R_j
\in \mathcal {R}_1} \frac{|h_{R_j,E_i}|^2}{d_{R_j,E_i}^\alpha}} \geq
\gamma_E\bigg|G_2^{(i)}\right)\left(1- \frac{1}{2} \pi {d_0}^2\right)\\
\end{aligned}
\end{equation*}

of which

\begin{equation*}
\begin{aligned}
&P\left(\frac{\frac{|h_{S,E_i}|^2}{d_{S,E_i}^\alpha}}{\sum_{R_j \in
\mathcal {R}_1} \frac{|h_{R_j,E_i}|^2}{d_{R_j,E_i}^\alpha}} \geq
\gamma_E\bigg|G_2^{(i)}\right)\\
&\ \ \leq P\left(\frac{|h_{S,E_i}|^2 {d_0}^{-\alpha}}{\Gamma
\int_0^1\int_0^1\frac{1}{\left[\left(x - x_{E_i}\right)^2 + \left(y
- y_{E_i}\right)^2\right]^{\frac{\alpha}{2}}}dxdy}\geq
\gamma_E\bigg|G_2^{(i)}\right)
\end{aligned}
\end{equation*}

where $\left(x_{E_i}, y_{E_i}\right)$ is the coordinate of the
eavesdropper $E_i$. $\Gamma$ is the sum of
$\left(n-1\right)\left(1-e^{-\tau}\right)$ independent exponential
random variables.

From Fig 2 we know that the strongest interference at eavesdropper
$E_i$ happens when $E_i$ is located in the point $(0, 0)$, while the
smallest interference at $E_i$ happens it is located at four corners
of the network region. By considering the smallest interference at
eavesdroppers, we then have

\begin{align*}
&P\left(C_{S,E_i} \geq \gamma_E\bigg|G_2^{(i)}\right)\\
&\ \ \ \ \ \ \leq P\left(\frac{|h_{S,E_i}|^2 {d_0}^{-\alpha}}{\Gamma
\psi }\geq
\gamma_E\right)\\
&\ \ \ \ \ \ = P\left(|h_{S,E_i}|^2 \geq \Gamma \gamma_E \cdot \psi
\cdot {d_0}^{\alpha} \right)
\end{align*}

here

$$\psi =
\int_{-0.5}^{0.5}\int_{-0.5}^{0.5}\frac{1}{\left[\left(x-0.5\right)^2
+ \left(y-0.5\right)^2\right]^{\frac{\alpha}{2}}}dxdy$$

Based on the Markov inequality,

\begin{align*}
& P\left(C_{S,E_i} \geq \gamma_E\bigg|G_2^{(i)}\right)\\
& \ \ \ \ \ \leq E_\Gamma \left[e^{-\Gamma \gamma_E \psi
{d_0}^{\alpha}}\right]\\
& \ \ \ \ \ = \left(\frac{1}{1 + \gamma_E \psi
{d_0}^{\alpha}}\right)^{\left(n-1\right)\left(1-e^{-\tau}\right)}\\
\end{align*}

Then, we have

\begin{equation}
\begin{aligned}
& P\left(C_{S,E_i} \geq \gamma_E\right) \\
&\leq \frac{1}{2} \pi {d_0}^2 +\left(\frac{1}{1 + \gamma_E \psi
{d_0}^{\alpha}}\right)^{\left(n-1\right)\left(1-e^{-\tau}\right)}\left(1- \frac{1}{2} \pi {d_0}^2\right)\\
\end{aligned}
\end{equation}

Employee the same method, we have

\begin{equation}
\begin{aligned}
& P\left(C_{R_{j^\ast},E_i} \geq \gamma_E\right) \\
&\leq \pi {d_0}^2 + \left(\frac{1}{1 + \gamma_E \psi
{d_0}^{\alpha}}\right)^{\left(n-1\right)\left(1-e^{-\tau}\right)}\left(1- \pi {d_0}^2\right)\\
\end{aligned}
\end{equation}

Notice that

\begin{equation}
\begin{aligned}
&\frac{1}{2} \pi {d_0}^2 +\left(\frac{1}{1 + \gamma_E \psi
{d_0}^{\alpha}}\right)^{\left(n-1\right)\left(1-e^{-\tau}\right)}\left(1- \frac{1}{2} \pi {d_0}^2\right)\\
&= \pi {d_0}^2 + \left(\frac{1}{1 + \gamma_E \psi
{d_0}^{\alpha}}\right)^{\left(n-1\right)\left(1-e^{-\tau}\right)}\left(1- \pi {d_0}^2\right)\\
&\ \ \ \ -\frac{1}{2} \pi {d_0}^2 \left[1-\left(\frac{1}{1 +
\gamma_E \psi
{d_0}^{\alpha}}\right)^{\left(n-1\right)\left(1-e^{-\tau}\right)}\right]\\
& \leq \pi {d_0}^2 + \left(\frac{1}{1 + \gamma_E \psi
{d_0}^{\alpha}}\right)^{\left(n-1\right)\left(1-e^{-\tau}\right)}\left(1-
\pi {d_0}^2\right)
\end{aligned}
\end{equation}

From formula (36), (37) and (38), we can get

\begin{equation}
\begin{aligned}
&P\left(O_{S \rightarrow R_{j^\ast}}^{(S)}\right) \leq P\left(O_{R_{j^\ast} \rightarrow D}^{(S)}\right)  \\
&\leq m \left[\pi {d_0}^2 + \left(\frac{1}{1 + \gamma_E \psi
{d_0}^{\alpha}}\right)^{\left(n-1\right)\left(1-e^{-\tau}\right)}\left(1- \pi {d_0}^2\right)\right]\\
\end{aligned}
\end{equation}

Substituting formula (39) into formula (2), we have

\begin{equation}
\begin{aligned}
&P_{out}^{\left(S\right)} \leq \\
&2 m \left[\pi {d_0}^2 + \left(\frac{1}{1 + \gamma_E \psi
{d_0}^{\alpha}}\right)^{\left(n-1\right)\left(1-e^{-\tau}\right)}\left(1- \pi {d_0}^2\right)\right]\\
&-\left[m \left(\pi {d_0}^2 + \left(\frac{1}{1 + \gamma_E \psi
{d_0}^{\alpha}}\right)^{\left(n-1\right)\left(1-e^{-\tau}\right)}\left(1-
\pi {d_0}^2\right)\right)\right]^2
\end{aligned}
\end{equation}

\end{proof}

\section{The Distribution Function and Probability
Density of $H = min\left(|h_{S,R_{j^\ast}}|^2,
|h_{D,R_{j^\ast}}|^2\right)$}

Let the random variable $H = min\left(|h_{S,R_{j^\ast}}|^2,
|h_{D,R_{j^\ast}}|^2\right)$. The node $R_{j^\ast}$ is randomly
selected from the relay selection set consisting of system nodes
with the first $k$ large $min\left(|h_{S,R_j}|^2,
|h_{D,R_j}|^2\right)$, $j=1,2,\cdots,n$. The distribution function
and probability density of $H$ is given by

\begin{equation*}
F_{H}\left(x\right)=\begin{cases} \frac{1}{k}\sum \limits_{j=1}^k
\bigg[ \sum \limits_{i=n-j+1}^n
\binom{n}{i}\\
\ \ \ \ \ \ \
\left[1-e^{-2x}\right]^{i}\left[e^{-2x}\right]^{n-i}\bigg] \ \ \ &
\text{$x > 0$}\\
0 \ \ \ &
\text{$x \leq 0$}\\
\end{cases}
\end{equation*}

and

\begin{equation*}
f_{H}\left(x\right) =\begin{cases} \frac{1}{k}\sum \limits_{j=1}^k
\bigg[
\frac{n!}{(j-1)!(n-j)!}\cdot\\
\ \
\left[1-e^{-2x}\right]^{n-j}\left[e^{-2x}\right]^{j-1}\left[2e^{-2x}\right]\bigg]
\ \ \ &
\text{$x > 0$}\\
0 \ \ \ &
\text{$x \leq 0$}\\
\end{cases}
\end{equation*}

\begin{proof}

Because the random variable $H = min\left(|h_{S,R_{j^\ast}}|^2,
|h_{D,R_{j^\ast}}|^2\right)$ is the random selection relay from the
first $k$ large random variable $min\left(|h_{S,R_j}|^2,
|h_{D,R_j}|^2\right)$, $j=1,2,\cdots,n$. From Appendix D,

\begin{align*}
& F_{H}\left(x\right) = \frac{1}{k}\sum \limits_{j=1}^k F_{H_j^l}(x)
\end{align*}

\begin{align*}
& f_{H}\left(x\right) = \frac{1}{k}\sum \limits_{j=1}^k f_{H_j^l}(x)
\end{align*}

According to Appendix E, we have

\begin{equation*}
F_{H}\left(x\right)=\begin{cases} \frac{1}{k}\sum \limits_{j=1}^k
\bigg[ \sum \limits_{i=n-j+1}^n
\binom{n}{i}\\
\ \ \ \ \ \ \
\left[1-e^{-2x}\right]^{i}\left[e^{-2x}\right]^{n-i}\bigg] \ \ \ &
\text{$x > 0$}\\
0 \ \ \ &
\text{$x \leq 0$}\\
\end{cases}
\end{equation*}

and

\begin{equation*}
f_{H}\left(x\right) =\begin{cases} \frac{1}{k}\sum \limits_{j=1}^k
\bigg[
\frac{n!}{(j-1)!(n-j)!}\cdot\\
\ \ \
\left[1-e^{-2x}\right]^{n-j}\left[e^{-2x}\right]^{j-1}\left[2e^{-2x}\right]\bigg]
\ \ \ &
\text{$x > 0$}\\
0 \ \ \ &
\text{$x \leq 0$}\\
\end{cases}
\end{equation*}

\end{proof}

\section{The Randomly Selected Variable from the Random Variable Set}

Let $X_1, \cdots ,X_n$ be continuous random variables, with density
$f_{X_1}(x),\cdots ,f_{X_n}(x)$ and distribution function
$F_{X_1}(x),\cdots ,F_{X_n}(x)$. The random variable, indexed by
$Y$, is selected randomly from $X_1, \cdots ,X_n$. The distribution
function and probability density of $Y$ is given by

\begin{align*}
& F_{Y}\left(y\right) = \frac{1}{n}\sum_{i=1}^n F_{X_i}(y)
\end{align*}

\begin{align*}
& f_{Y}\left(y\right) = \frac{1}{n}\sum_{i=1}^n f_{X_i}(y)
\end{align*}

\begin{proof}

We assume the $s$-th random variable is selected as $Y$, $P(s=i) =
\frac{1}{n}, i= 1, \cdots ,n$. Then we have

\begin{align*}
& F_{Y}\left(y\right) = P\left(Y \leq y\right)\\
&\ \ \ \ \ \ \ \ = \sum_{i=1}^n P\left(X_s \leq y|s =
i\right)P\left(s = i\right)\\
&\ \ \ \ \ \ \ \ = \sum_{i=1}^n \frac{1}{n}P\left(X_s \leq y|s =
i\right)\\
&\ \ \ \ \ \ \ \ = \sum_{i=1}^n \frac{1}{n}P\left(X_i \leq y\right)\\
&\ \ \ \ \ \ \ \ = \frac{1}{n}\sum_{i=1}^n F_{X_i}(y)
\end{align*}

\begin{align*}
& f_{Y}\left(y\right) = F_{Y}^{\prime}\left(y\right)\\
&\ \ \ \ \ \ \ \ = \frac{1}{n}\sum_{i=1}^n F_{X_i}^{\prime}(y)\\
&\ \ \ \ \ \ \ \ = \frac{1}{n}\sum_{i=1}^n f_{X_i}(y)
\end{align*}

\end{proof}

\section{The Distribution Function and Probability
Density of the $k$-th Largest Random Variable}

The $|h_{A,B}|^2$ is path-loss between any node $A$ and $B$ with the
Rayleigh fading, and is exponentially distributed with
$E\left[|h_{A,B}|^2\right] = 1$. The $min\left(|h_{S,R_j}|^2,
|h_{D,R_j}|^2\right)$, $j=1,2,\cdots,n$, are $n$ random variables in
which the $j$-th largest random variable is denoted by $H_j^l$. The
distribution function and probability density of the random variable
$H_j^l$, $j=1,2,\cdots,n$, are given by

\begin{equation*}
F_{H_j^l}\left(x\right)=\begin{cases} \sum \limits_{i=n-j+1}^n
\binom{n}{i}\left[1-e^{-2x}\right]^{i}\left[e^{-2x}\right]^{n-i} \ \
\ &
\text{$x > 0$}\\
0 \ \ \ &
\text{$x \leq 0$}\\
\end{cases}
\end{equation*}

\begin{equation*}
f_{H_j^l}\left(x\right) =\begin{cases}
\frac{n!}{(j-1)!(n-j)!}\cdot\\
\left[1-e^{-2x}\right]^{n-j}\left[e^{-2x}\right]^{j-1}\left[2e^{-2x}\right]
\ \ \ &
\text{$x > 0$}\\
0 \ \ \ &
\text{$x \leq 0$}\\
\end{cases}
\end{equation*}

\begin{proof}

Because the $|h_{A,B}|^2$ is exponentially distributed with
$E\left[|h_{A,B}|^2\right]=1$ between any node $A$ and $B$,
according to order statistics in \cite{IEEEhowto:David}, we can get
the distribution function of the $min\left(|h_{S,R_j}|^2,
|h_{D,R_j}|^2\right)$ for each relay $R_j, j= 1, 2, \cdots, n$,
indexed by $H_j$, as following,

\begin{equation*}
f_{H_j}(x)=\begin{cases} 2e^{-2x} \ \ \ &
\text{$x > 0$}\\
0 \ \ \ &
\text{$x \leq 0$}\\
\end{cases}
\end{equation*}

\begin{equation*}
F_{H_j}(x)=\begin{cases} 1-e^{-2x} \ \ \ &
\text{$x > 0$}\\
0 \ \ \ &
\text{$x \leq 0$}\\
\end{cases}
\end{equation*}

According to order statistics in \cite{IEEEhowto:David}, The
distribution function and probability density of the $j$-th smallest
in $H_j$, $j=1,2,\cdots,n$, indexed by $H_j^s$, are given by

\begin{equation*}
F_{H_j^s}\left(x\right) = \begin{cases} \sum \limits_{i=j}^n
\binom{n}{i}\left[1-e^{-2x}\right]^{i}\left[e^{-2x}\right]^{n-i} \ \
\ &
\text{$x > 0$}\\
0 \ \ \ &
\text{$x \leq 0$}\\
\end{cases}
\end{equation*}

\begin{equation*}
f_{H_j^s}\left(x\right) = \begin{cases}
\frac{n!}{(j-1)!(n-j)!}\cdot\\
\left[1-e^{-2x}\right]^{j-1}\left[e^{-2x}\right]^{n-j}\left[2e^{-2x}\right]
\ \ &
\text{$x > 0$}\\
0 \ \ \ &
\text{$x \leq 0$}\\
\end{cases}
\end{equation*}

Since the $j$-th largest, indexed by $H_j^l$, is equal to the
$\left(n-j+1\right)$-th smallest in $H_j$, $j=1,2,\cdots,n$, we
should have

\begin{align*}
& F_{H_j^l}\left(x\right) = F_{H_{n-j+1}^s}\left(x\right)\\
&\ \ \ \ = \begin{cases} \sum \limits_{i=n-j+1}^n
\binom{n}{i}\left[1-e^{-2x}\right]^{i}\left[e^{-2x}\right]^{n-i} \ \
&
\text{$x > 0$}\\
0 \ \ \ &
\text{$x \leq 0$}\\
\end{cases}\\
\end{align*}

\begin{align*}
&f_{H_j^l}\left(x\right) = f_{H_{n-j+1}^s}\left(x\right)\\
&\ \ \ \ = \begin{cases} \frac{n!}{(j-1)!(n-j)!}\cdot\\
\left[1-e^{-2x}\right]^{n-j}\left[e^{-2x}\right]^{j-1}\left[2e^{-2x}\right]
\ \ &
\text{$x > 0$}\\
0 \ \ \ &
\text{$x \leq 0$}\\
\end{cases}\\
\end{align*}

\end{proof}



\ifCLASSOPTIONcaptionsoff
\newpage
\fi

\end{document}